\newcommand{\CLASSINPUTinnersidemargin}{18mm}
\newcommand{\CLASSINPUToutersidemargin}{12mm}
\newcommand{\CLASSINPUTtoptextmargin}{20mm}
\newcommand{\CLASSINPUTbottomtextmargin}{25mm}
\documentclass[10pt,conference,a4paper]{IEEEtran}

\usepackage{times}

\usepackage{graphicx}

\usepackage{amsmath}
\usepackage{times}
\usepackage{graphicx}
\usepackage{amsfonts}
\usepackage{amssymb}
\usepackage{cite}
\usepackage{subfigure}
\usepackage{url}
\usepackage{color}
\usepackage{upgreek}

\DeclareGraphicsExtensions{.png,.eps,.ps,.pdf}

\begin{document}

\title{Simulador electromagnético eficiente para el diseño de rejillas de difracción débiles en guías dieléctricas}


\author{
\authorblockN{Elio Godoy-Lorite, Laureano Moreno-Pozas, José Manuel Luque-González, Ana Sánchez-Ramírez,\\ Robert Halir, Alejandro Ortega-Moñux, J. Gonzalo Wangüemert-Pérez, Iñigo Molina-Fernández}
\authorblockA{\{eliogl, lmp, jmlg, asr\}@ic.uma.es; rhalir@uma.es; \{aom, gonzalo\}@ic.uma.es; imolina@uma.es.}
\authorblockA{Telecommunication Research Institute (TELMA), Universidad de M\'alaga, Louis Pasteur 35, 29010 Málaga, Spain.}

}

\maketitle

\begin{abstract}
The significant growth of free-space optic communications and Light Detection and Ranging (LiDAR) is demanding gratings that emit highly collimated beams, i.e. with Rayleigh ranges of millimeters or even centimeters. Hence, weak-strength gratings, which radiate little amount of power per unit length, are needed. The main purpose of this work is to propose an efficient and accurate simulation tool to accelerate the design of weak-strength gratings required for these applications. To achieve this, we propose a simulator based on the classical perturbation method, which takes a circuit approach to the electromagnetic problem. Comparison with results obtained with a rigorous 2D full wave electromagnetic simulator (FEXEN) shows very good agreement with the advantage of decreasing simulation times by up to a factor $\times 16$.

\end{abstract}

\section{Introducción}

La fotónica de silicio (SiPh) se encuentra en la vanguardia de las futuras comunicaciones ópticas de alta velocidad, gracias a que permite avanzar hacia la integración monolítica de transceptores completos en una tecnología de fabricación compatible con la usada por la industria microelectrónica CMOS, lo que posibilita escalados de grandes volúmenes con costes reducidos \cite{roadmapping}. En la gran cantidad de aplicaciones de SiPh, que van desde la óptica de espacio libre \cite{sota}, el LiDAR \cite{lidar} o los biosensores \cite{Ana}, las rejillas de difracción son dispositivos clave. Aunque la aplicación más habitual de las rejillas de difracción ha sido como interfaces entre la fibra óptica y el chip  (acopladores chip-fibra) \cite{chipFibra}, en algunas aplicaciones, como por ejemplo el sensado remoto, se requieren rejillas de gran apertura que emitan haces muy colimados, cuyas distancias de Rayleigh superen los centímetros \cite{ropp2021meta}, es decir, rejillas con baja fuerza de radiación. Debido a la gran importancia de estos dispositivos, se propone el desarrollo de un simulador electromagnético optimizado para el diseño de estas rejillas de difracción. Aprovechando que las rejillas que se pretenden diseñar son débiles, es razonable suponer que los métodos perturbacionales (MP) pueden ofrecer simultáneamente una buena precisión y una alta eficiencia computacional. Por ello, el simulador desarrollado se basa en el método perturbacional propuesto por Tamir et al. \cite{tamir} que se explica en la Sección \ref{sec:metodo}. En la Sección \ref{sec:Simulador} se describen sus funcionalidades, y, finalmente, su potencial se ilustra en la Sección \ref{sec:Disenio}. En ella se realiza un diseño de una rejilla de difracción de baja fuerza de radiación con una longitud total de $0.4\ \rm{mm}$, que emite haces con distancia de Rayleigh de más de $4\ \rm{cm}$ a una longitud de onda en el vacío $\lambda_0=1.55\ \upmu\rm m$. Se comparan los resultados de simulación con el simulador electromagnético 2D FEXEN \cite{FEXEN}, un simulador desarrollado y empleado ampliamente en el grupo de investigación en el cual se ha desarrollado este trabajo. Los tiempos de cálculo son muy favorables si se comparan con los requeridos por FEXEN, que emplea el método de expansión modal de Fourier, y que ya es más rápido, para el analisis de rejillas de difracción, que el método de diferencias finitas en el dominio del tiempo FDTD \cite{FEXEN}. En esta comparación, también se demuestra que el simulador aquí presentado no disminuye la precisión en el cómputo de características principales de las rejillas de difracción, como son el campo cercano generado, la direccionalidad y la dirección y fuerza de radiación.

\section{El método perturbacional}
\label{sec:metodo}

\begin{figure}[t!]
\centering
\includegraphics[width=0.9\columnwidth]{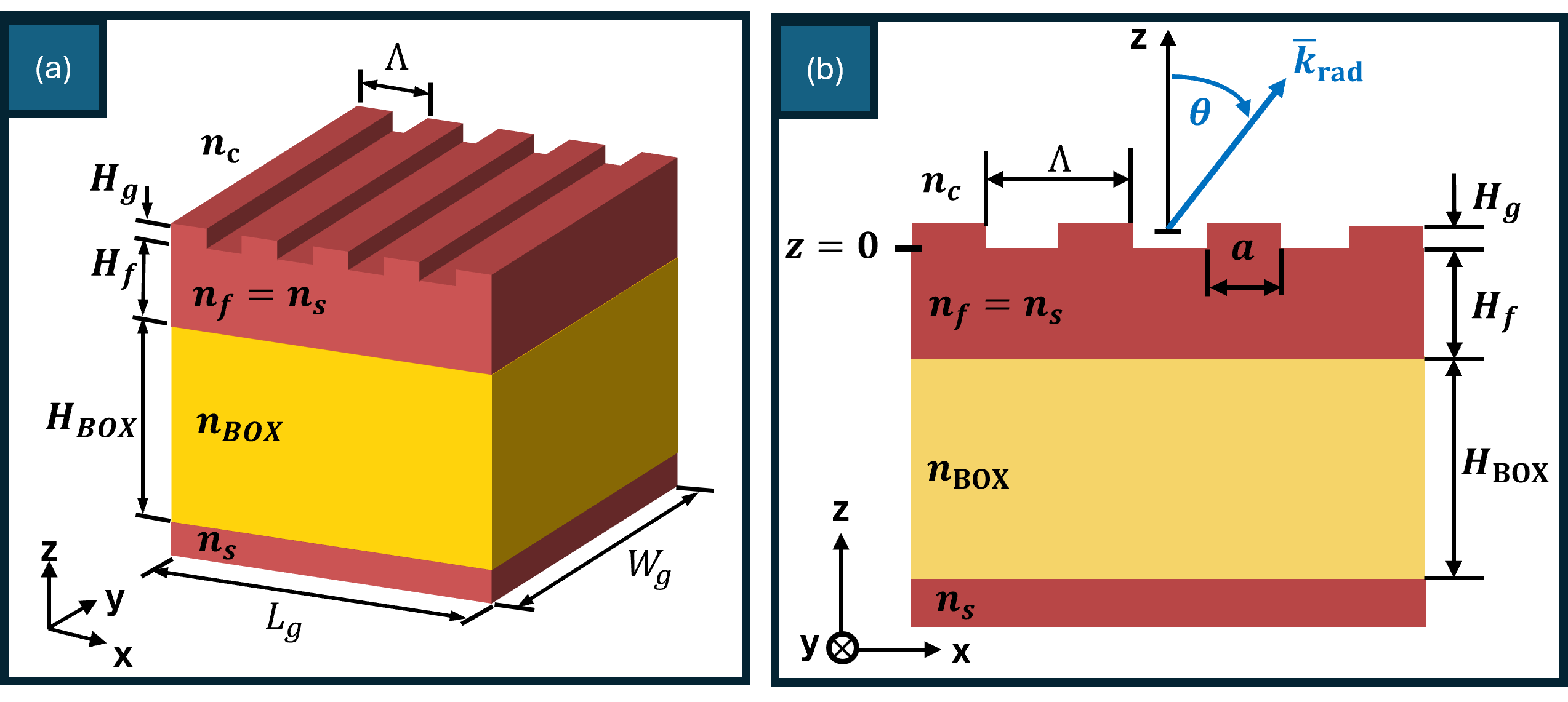}
\caption{Esquemático de una rejilla de difracción en SOI. (a) Estructura 3D. (b) Corte longitudinal (plano XZ) de la estructura. Por claridad, no se ha coloreado la cubierta.}
\label{fig:Grating2D}
\end{figure}

El MP que se ha implementado en este trabajo fue desarrollado y aplicado al análisis de rejillas de difracción en guías dieléctricas por Tamir et al. en  \cite{tamir}. Sin embargo, hasta donde los autores conocen, es la primera vez que se prueba en una tecnología de alto contraste como es la tecnología de Silicio sobre aislante ("Silicon-On-Insulator", SOI), empleando la plataforma de fabricación estándar con $H_f+H_g=220\ \rm{nm}$ y $H_{\rm{BOX}}=2\ \upmu\rm m$, la cual se ilustra en la Fig.~\ref{fig:Grating2D}(a). El estándar en plataforma SOI presenta unos índices de refracción en el núcleo y el sustrato iguales $n_f=n_s = 3.476$, y también para el óxido enterrado (BOX) y la cubierta, $n_{\rm{BOX}}=n_c=1.444$. El método se puede emplear para realizar el análisis perturbacional de cualquier estructura laminar dieléctrica donde una de sus láminas tiene una constante dieléctrica periódica arbitraria que se considera como una perturbación débil en la estructura. En el caso de este trabajo, se va a particularizar este método para el caso de la rejilla de difracción de la Fig.~\ref{fig:Grating2D}(a), donde \(H_g \ll H_f\) (perturbación débil). Aquí se supone que la guía de láminas de la Fig.~\ref{fig:Grating2D} es suficientemente ancha ($W_g>10\ \upmu\rm m$), y la excitación se supone invariante en dirección \(y\). Por ello, el problema relativo al estudio de la propagación de los campos por la estructura en dirección \(x\), así como el acoplo de parte de la potencia de dichos campos a una onda plana cuyo vector de onda es $\bar{k}_{\rm{rad}}$, se reduce a la resolución de un problema 2D en el plano XZ, como el de la Fig.\ref{fig:Grating2D}(b). Así, en la estructura de la Fig.~\ref{fig:Grating2D}(b), se puede expresar la constante dieléctrica como una función de $x$ y de $z$ de la forma siguiente:
\begin{equation}
	\epsilon(x,z) = \epsilon_h(z) + \epsilon_p(x),
	\label{eq:epsilonEstructura}
\end{equation}
donde \(\epsilon_h(z)\) agrupa la media volumétrica de la constante dieléctrica de cada capa en un corte $z$, es decir,
\begin{equation}
    \epsilon_h(z) = \frac{1}{\Lambda}\int_{0}^{\Lambda}\epsilon(x,z)dx,
    \label{eq:epsilonH}
\end{equation}
y el término \(\epsilon_p(x)\) introduce la variación periódica de la constante dieléctrica en torno a esta media, con lo que será distinta de cero solo cuando \(z\in[0,H_g]\).  Esta división en dos partes del problema va a facilitar descomponer las ecuaciones de Maxwell en dos conjuntos de ecuaciones: las asociadas a las componentes de campo de la estructura homogénea, sin perturbar, ($\bar{E}_h$ para el campo eléctrico y $\bar{H}_h$ para el magnético) y las asociadas a las componentes de campo de la estructura perturbada ($\bar{E}_p$ para el campo eléctrico y $\bar{H}_p$ para el magnético), donde $\bar E = \bar{E}_h+\bar{E}_p$ y $\bar{H}=\bar{H}_h+\bar{H}_p$.
Introduciendo esta separación de los campos junto con (\ref{eq:epsilonEstructura}) en las ecuaciones de Maxwell, se llega a las siguientes ecuaciones asociadas a la estructura perturbada: 
\begin{equation}
    \left\{\begin{matrix}
   \nabla\times \bar{E}_p=-j\omega\mu_0\bar{H}_p \\
    \nabla\times \bar{H}_p=j\omega\epsilon_0\left(\epsilon_h\bar{E}_p+\epsilon_p\bar{E}_h+\epsilon_p\bar{E}_p\right)
\end{matrix}\right..\label{eq:RotEHp}
\end{equation}

Teniendo en cuenta que se está trabajando con perturbaciones débiles, el término $\epsilon_p\bar{E}_p$ de la ecuación (\ref{eq:RotEHp}) se considera despreciable frente al resto. De esta forma, si se desarrollan los rotacionales, se realiza el cambio de variable de la Tabla \ref{tab:CamposPerVI} y se expande en series de Fourier la constante dieléctrica $\epsilon_p$ de la ecuación (\ref{eq:epsilonEstructura}), se obtienen $n$ grupos de ecuaciones, uno por cada armónico $\epsilon_p^n$, de la forma 
\begin{equation}
\left\{\begin{matrix}
    \frac{\partial V_p^{n} }{\partial z}=-jk_{z,h}^{n}Z_h^{n}I_p^{n}+v^{n}  \\
    \frac{\partial I_p^{n} }{\partial z}=-j\frac{k_{z,h}^{n}}{Z_h^{n}}V_p^{n}+i^{n}
\end{matrix}\right.,\label{eq:derivVpz}
\end{equation}
donde $n$ es el orden del armónico de $\epsilon_p$. Identificando estas ecuaciones en (\ref{eq:derivVpz}) con una versión modificada de las ecuaciones del telegrafista de una línea de transmisión \cite{ cheng}, se puede hacer una analogía entre el problema electromagnético y un problema de líneas de transmisión donde se incorporan una serie de fuentes de tensión y corriente cuyas densidades son $v^{n}$, $i^{n}$, con unidades de $\rm{V/m}$ y $\rm{A/m}$, respectivamente.
\begin{table}[h]
\renewcommand{\arraystretch}{1.3}
\caption{Ecuaciones para realizar el análisis de rejillas de difracción mediante líneas de transmisión. $k_{x}^{n}$ calculada con (\ref{eq:ecuacionGrating}).}
\begin{center}
\begin{tabular}{ccc}
TE         & TM        \\ \hline
\rule{0pt}{1\normalbaselineskip}\(\bar{H}_{xp}=-\sum_{n}I_p^{n}(z)\)          &  \(\bar{E}_{xp}=\sum_{n}V_p^{n}(z)\)         \\[0.12cm]
\(\bar{E}_{yp}=\sum_{n}V_p^{n}(z)\)           & \(\bar{H}_{yp}=\sum_{n}I_p^{n}(z)\)          \\[0.12cm]
\(\bar{H}_{zp}=\frac{k_{x}^{n}}{\omega\mu_0}V_p^{n}(z)\) & \(\bar{E}_{zp}=-\sum_{n}\frac{k_{x}^{n}}{\omega\epsilon_0\epsilon_h}I_p^{n}(z)\)\\[0.2cm]\hline
\rule{0pt}{1.2\normalbaselineskip}
\(v^{n}(z)=0\)         & \(v^{n}(z)=-j\frac{\beta_{\rm{eff}}k_{x}^{n}\epsilon_{p}^{n}(z)}{\omega\epsilon_0\epsilon_g^{2}}I_{g,h}(z)\)         \\[0.12cm]
\(i^{n}(z)=j\omega\epsilon_0\epsilon_{p}^{n}(z)V_{g,h}(z)\)           & \(i^{n}(z)=j\omega\epsilon_0\epsilon_{p}^{n}(z)V_{g,h}(z)\)        \\[0.1cm]\hline
\rule{0pt}{1\normalbaselineskip}
\(Z_h^n=\frac{1}{Y_h^n}=\frac{k_{z,h}^{n}}{\omega\mu_0}\)           &\(Z_h^n=\frac{1}{Y_h^n}=\frac{\omega\epsilon_0\epsilon_h }{k_{z,h}^{n}}\)            \\[0.15cm]
\multicolumn{2}{c}{\(k_{z,h}^{n}=(k_0^{2}\epsilon_h-(k_{x}^{n})^{2})^{1/2}\)} 
\\ [0.1cm]\hline
\end{tabular}
\end{center}
\label{tab:CamposPerVI}
\end{table}
\begin{figure}[t]
\centering
\includegraphics[width=0.9\columnwidth]{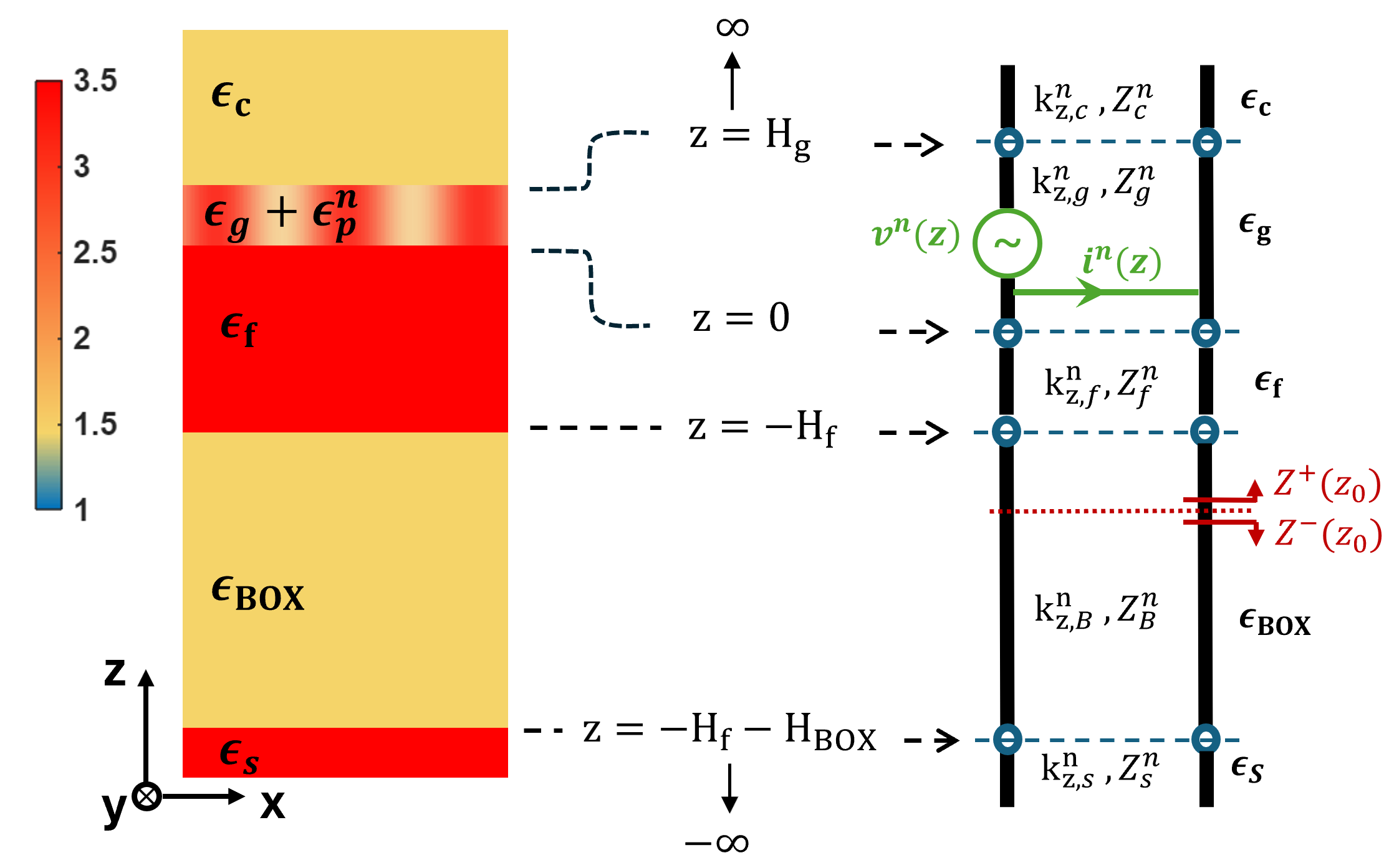}
\caption{Modelo de líneas de transmisión equivalente a la estructura dieléctrica donde la lámina periódica original se sustituye por su armónico $n$-ésimo.}
\label{fig:ModeloLinea}
\end{figure}

Para ilustrar esta analogía, se presenta en la Fig. \ref{fig:ModeloLinea} la estructura correspondiente al armónico $n$-ésimo de la descomposición en serie de Fourier del término $\epsilon_p$. Cada lámina homogénea de la estructura original, se modela como una línea de transmisión de longitud igual a la altura de la lámina original y con constantes de propagación (en la figura $k_{z,h}^n$) e impedancias características ($Z_h^n$) distintas para cada armónico. Las fuentes de corriente y tensión aparecerían exclusivamente en la capa de espesor $H_g$. Cabe destacar que, aunque en la ilustración solo se ha presentado una fuente de tensión y de corriente, estas representan un continuo de fuentes de tensión y corriente a lo largo de la línea de longitud $H_g$. Para $n=0$, no existen dichas fuentes ya que $\epsilon_p^n(z)=0$.

Como estas fuentes de tensión y corriente solo dependen de las ondas de tensión y corriente soportadas por la estructura cuando $n=0$, se tendrá que resolver primero el problema para $n=0$ (problema sin perturbar), donde la capa de espesor $H_g$ se sustituye por una capa homogénea cuya constante dieléctrica es $\epsilon_g$ calculada a partir de la ecuación (\ref{eq:epsilonH}). Después se resolverá el problema para cada uno de los armónicos, (problema perturbado) truncando el problema para un número de armónicos suficientemente elevado.

Para resolver la estructura de líneas de transmisión para $n=0$, se empleará la condición de resonancia transversal $\rm{Im}\left \{  Z^{+}(z_0) + Z^{-}(z_0)\right \} = 0$,
donde $Z^{-}(z_0)$ y $Z^{+}(z_0)$ son las impedancias vistas en sentido opuesto en un plano $z_0$ cualquiera de la línea, dando como resultado el índice efectivo del modo guiado $n_{\rm{eff}}$, ya sea un modo TE o TM. Una vez hallado este índice, se impone una determinada tensión en un plano cualquiera de la línea y, propagándola, se obtienen las ondas de tensión y corriente soportadas por la línea. Para $|n|\geq1$ se utilizan los resultados hallados para $n=0$ para modificar las líneas de transmisión según la \mbox{Tabla \ref{tab:CamposPerVI}}.

Siguiendo este procedimiento, es posible calcular de forma eficiente, por un lado, el perfil del modo de la guía periódica, como suma de cada una de las contribuciones de cada armónico y, por otro lado, la fuerza de radiación de la estructura, su direccionalidad y el ángulo de radiación para cada uno de los órdenes de difracción. Dichos ángulos se determinan a partir de la conocida ecuación de la rejilla \cite{tamir}
\begin{equation}
    \frac{k_x^n}{k_0}= n_c \sin(\theta) =n_{\rm{eff}}+n\frac{\lambda_0}{\Lambda}, \label{eq:ecuacionGrating}
\end{equation}
donde $k_0=2\pi/\lambda_0$, $n<0$, siempre y cuando se cumpla la condición de radiación, es decir, $\left|n_{\rm{eff}}+n\frac{\lambda_0}{\Lambda} \right| \leq n_c$.

\section{Simulador basado en el MP}
\label{sec:Simulador}

\begin{figure}[t!]
\centering
\includegraphics[width=0.6\columnwidth]{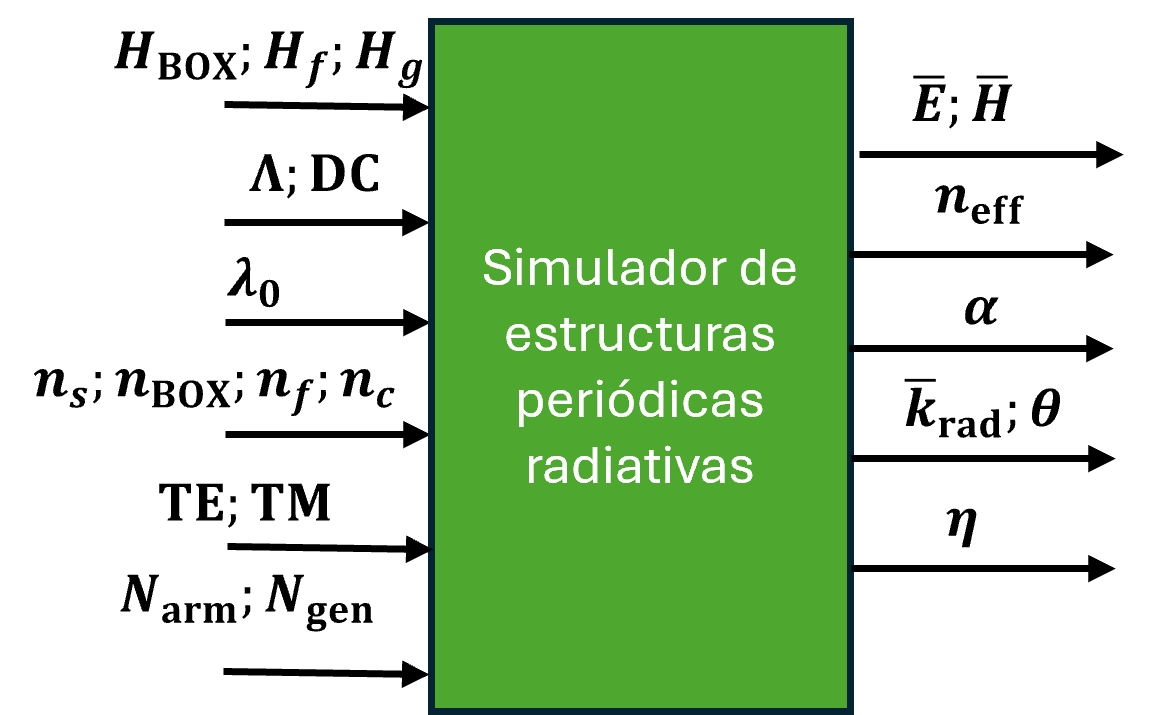}
\caption{Definición de entradas y salidas del simulador sistémico desarrollado en este trabajo.}
\label{fig:BloqueSistema}
\end{figure}

En la Fig. \ref{fig:BloqueSistema} se presenta el bloque sistémico del simulador basado en el MP. Este tiene como entradas la geometría de la estructura, tanto las alturas de cada lámina ($H_{\rm{BOX}}$, $H_f$, $H_g$), como el periodo ($\Lambda$) y el ciclo de trabajo ($DC=a/\Lambda$). También incluye los índices de refracción de los materiales ($n_s$, $n_{\rm{BOX}}$, $n_f$ y $n_c$), la longitud de onda ($\lambda_0$) y la polarización (TE o TM). Los parámetros de simulación son el número de armónicos al que se va a truncar el cálculo de la expansión de Fourier ($N_{\rm{arm}}$) y el número de fuentes $v^{n}$, $i^{n}$ por unidad de longitud ($N_{\rm{gen}}$).

Las salidas ofrecidas a partir de las entradas anteriores son: los perfiles de los campos totales de la estructura ($\bar{E}$ y $\bar{H}$), el índice efectivo de los modos soportados ($n_{\rm{eff}}$), la fuerza de radiación asociada a cada modo, en forma de constante de atenuación del modo guiado ($\alpha$), la dirección de radiación de cada orden de difracción ($\bar{k}_{\rm{rad}}$) y la direccionalidad de la estructura ($\eta$), definida como la ratio de potencia que radia hacia arriba frente a la que se radia hacia arriba y hacia abajo. Una de las grandes ventajas de este simulador es que el cálculo de estas salidas se hace en un tiempo de cómputo considerablemente menor al de FEXEN, tal y como se muestra en Tabla \ref{tab:TiemposComputo}. En ésta se compara el tiempo empleado para el cálculo de la direccionalidad $\eta$ y de las curvas de la Fig. \ref{fig:GraficasJuntas}, las cuales representan los principales barridos que hay que realizar a la hora de diseñar una rejilla de difracción como la propuesta en la Fig.~\ref{fig:Grating2D}, con $H_f=190\ \rm{nm}$ y $H_g=30\ \rm{nm}$. Por un lado, de las gráficas se observa que los resultados arrojados por el simulador son consistentes con FEXEN y, además, el tiempo empleado para los barridos es considerablemente menor, siendo especialmente menor para el caso del cálculo de la direccionalidad de la estructura final. 
Esto es debido a que el simulador basado en MP realiza todos los cálculos analíticos mediante teoría de líneas de transmisión, y únicamente tiene que calcular la fracción de potencia entregada a la línea superior frente a la total aplicando la teoría de circuitos:
\begin{equation}
    \eta = \frac{\frac{1}{2}\sum_n|V_{c,p}^n|^{2}\rm{Re}\left \{  Y_{c}^n\right \} }{\frac{1}{2}\sum_n\left(|V_{c,p}^n|^{2}\rm{Re}\left \{  Y_{c}^n\right \}+|V_{s,p}^n|^{2}\rm{Re}\left \{  Y_{s}^n\right \} \right)}
\end{equation}
donde $Y_c^n$ es la admitancia de la línea para el caso del armónico $n$-ésimo y $V_{c,p}^n$ es la tensión, ambas en el plano $z=H_g$, mientras que $Y_s^n$ y $V_{s,p}^n$ son las del plano ${z=-H_f-H_{\rm{BOX}}}$. Sin embargo, FEXEN necesita monitorizar el campo para comparar la potencia entregada en diferentes planos, lo cual es costoso computacionalmente. En cuanto a la fuerza de radiación $\alpha$, puesto que ésta indica la fracción de potencia radiada por unidad de longitud, se calcula aplicando los mismos conceptos como:
\begin{equation}
    \alpha =\frac{\frac{1}{2}p_{\rm{rad}}}{P} = \frac{\frac{1}{2}\sum_n \left(|V_{c,p}^n|^{2}\rm{Re}\left \{  Y_{c}^n\right \}+|V_{s,p}^n|^{2}\rm{Re}\left \{  Y_{s}^n\right \}\right)}{\int_{-\infty}^{\infty}\rm{Re}\left\{\bar{E}_h\times \bar{H}_h^{*} \right\}\cdot \hat{x}dz },
\end{equation}
donde el numerador indica la potencia entregada a las líneas exteriores por unidad de longitud, y el denominador indica la potencia transmitida en la dirección de propagación.

\begin{table}
\renewcommand{\arraystretch}{1.3}
\caption{Tabla comparativa de los tiempos de simulación de FEXEN y el simulador basado en el MP. Se ha utilizado un HP Pavilion con procesador Intel Core  i7-8565U CPU 1.99 GHz con 8GB de RAM.}
\label{tab:TiemposComputo}
\begin{center}
\begin{tabular}{|c|c|c|}
\hline
 & FEXEN & MP\\
\hline
Fig. \ref{fig:GraficasJuntas} (a) & 18 min 25 s& 6 min 57 s\\
\hline
Fig. \ref{fig:GraficasJuntas} (b) & 30 min 32 s& 3 min 18 s\\
\hline
$\eta$ & 4 min 10 s& 15 s\\
\hline

\end{tabular}
\end{center}
\end{table}

\section{Ejemplo de diseño de una rejilla de difracción}
\label{sec:Disenio}

Debido al creciente interés por el diseño de antenas ópticas integradas altamente directivas, con un tamaño de apertura varios órdenes de magnitud superior a la longitud de onda \cite{ropp2021meta}, en esta sección se va a aprovechar la rapidez y precisión ofrecida por el simulador aquí presentado para realizar el diseño de una rejilla de difracción en la plataforma SOI, de alto contraste, estableciendo una longitud mínima de la rejilla de $L_g=100\ \upmu \rm m$. Este valor se considera de interés para el estado del arte de este tipo de antenas, pues aperturas de esta magnitud solo se han conseguido en plataformas de bajo contraste como la de nitruro de silicio ($n_{\rm{SiN}} = 2$)\cite{ropp2021meta}. El diseño se hará sobre la plataforma SOI estándar descrita en la Fig.~\ref{fig:Grating2D} ($H_f+H_g=220\ \rm{nm}$ y $H_{\rm{BOX}}=2\ \upmu\rm m$), para polarización TE y trabajando con una longitud de onda en el vacío de $1.55\ \upmu\rm m$. Aunque aquí se ha supuesto que la cubierta es de altura infinita, se incorporará una restricción en su altura en una versión futura del simulador.

\begin{figure}[t]
\centering
\includegraphics[width=1\columnwidth]{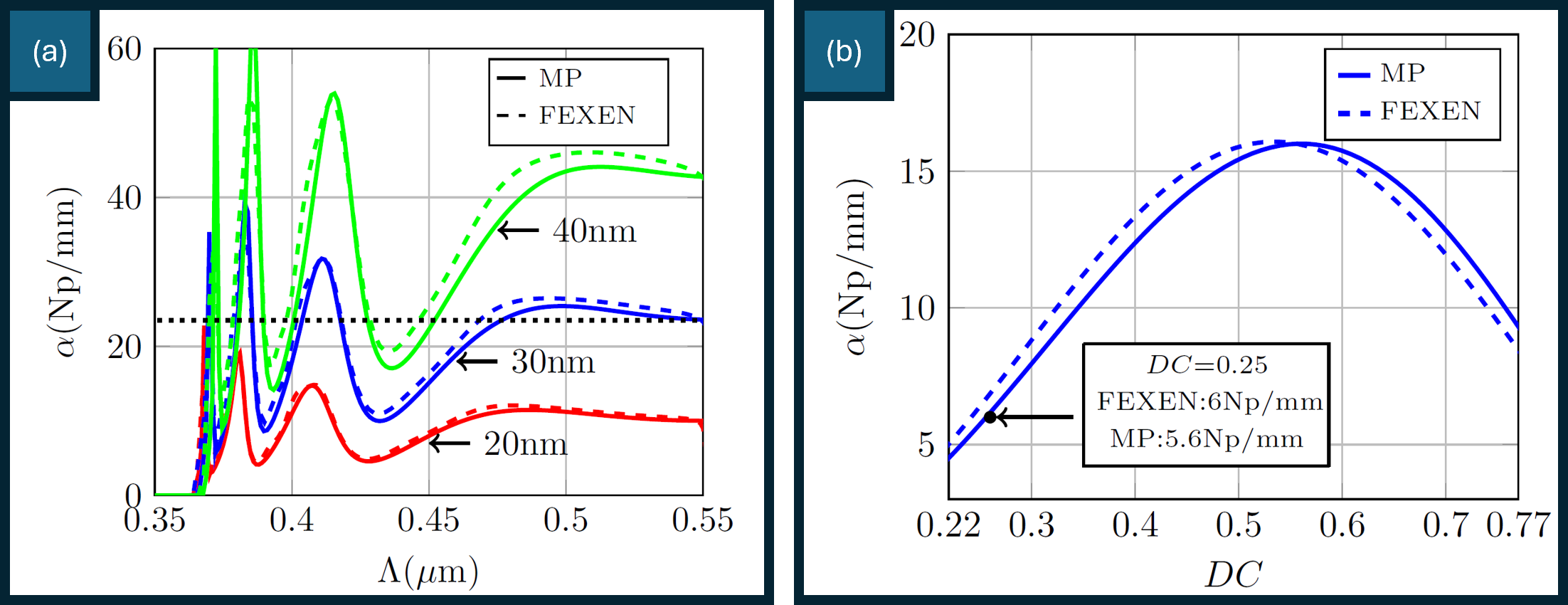}
\caption{(a) Fuerza de radiación en función del periodo de la rejilla con $DC=0.5\%$ para diferentes grabados $H_g$: $20\ \rm{nm}$ en rojo, $30\ \rm{nm}$ en azul y $40\ \rm{nm}$ en verde (b) Fuerza de radiación en función del ciclo de trabajo para $H_g=30\ \rm{nm}$.}
\label{fig:GraficasJuntas}
\end{figure}

\begin{figure}[t]
\centering
\includegraphics[width=1\columnwidth]{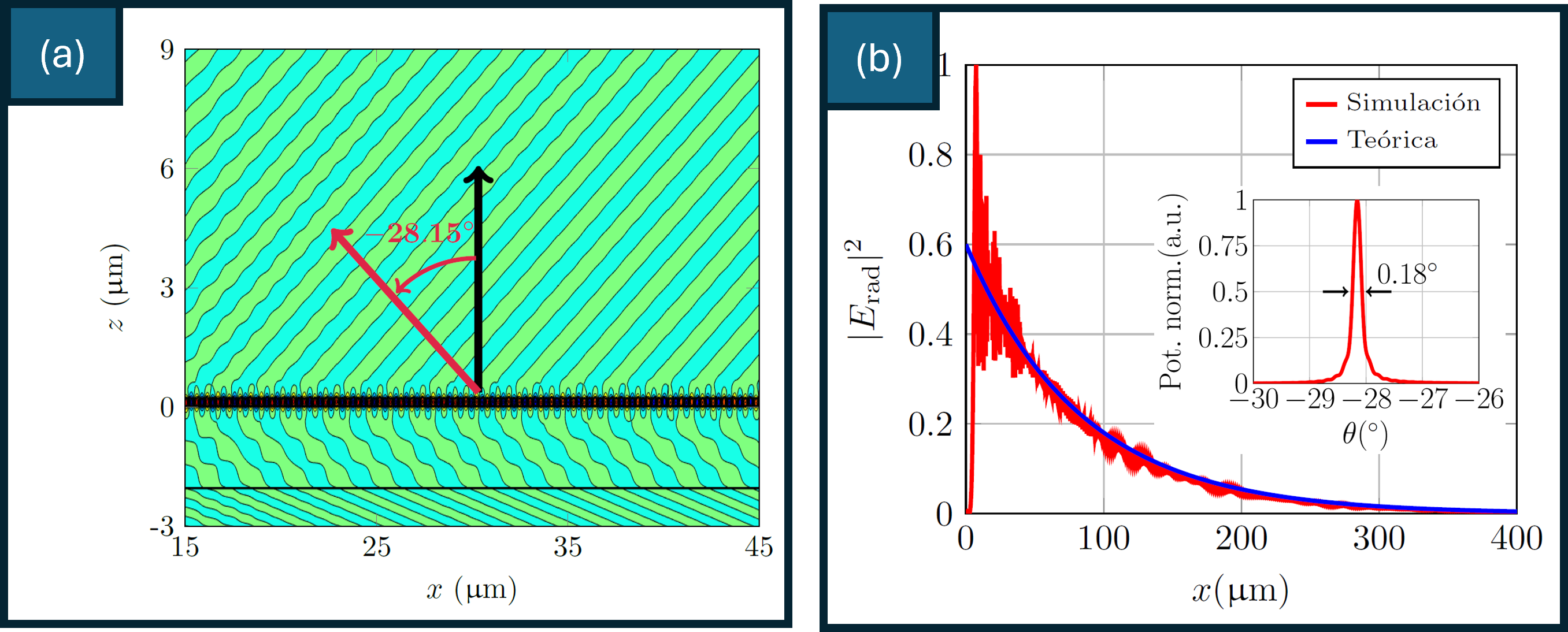}
\caption{(a) Simulación en FEXEN de la parte real del campo. (b) En rojo: intensidad del campo simulado con FEXEN en un corte longitudinal en la cubierta. En azul: exponencial teórica $|E_{\rm{rad}}|^{2}=0.6\exp(-2\alpha x)$, $\alpha = 5.6 \rm{Np/mm}$. En el recuadro interior: diagrama de radiación asociado.}
\label{fig:campo}
\end{figure}

La restricción sobre la longitud mínima de la rejilla impone una fuerza de radiación máxima de $\alpha=23\ \rm{Np/mm}$, la cual produce un haz con distancia de Rayleigh de $0.25\ \rm{cm}$. En primer lugar, se deberá elegir adecuadamente la altura del grabado $H_g$ para conseguir una fuerza de radiación cercana al valor deseado. Igualmente, se ha de elegir el periodo $\Lambda$ para trabajar en la primera zona de radiación y así evitar que varios órdenes de difracción cumplan con la condición de radiación, teniendo la precaución de no acercarse demasiado a la segunda zona de Bragg ($\Lambda < 0.55\ \upmu\rm m$) para evitar así reflexiones excesivas \cite{ Bragg}. Para ello se realiza el barrido de la Fig. \ref{fig:GraficasJuntas}(a), donde se ilustra la fuerza de radiación de una estructura con ciclo de trabajo del $50\%$, en función de diferentes pasos de grabado $H_g$: $20\ \rm{nm}$ en rojo, $30\ \rm{nm}$ en azul y $40\ \rm{nm}$ en verde. De las gráficas se descartan valores de periodo $\Lambda$ inferiores a $0.42\ \upmu\rm m$, al presentar una fuerte oscilación, que se traduce en una gran sensibilidad a errores de fabricación. Aquí se ha elegido $\Lambda=0.45\ \upmu\rm m$, ya que se haya cerca del mínimo de fuerza de radiación para los tres grabados con una variación lineal en torno al punto de diseño. Con este periodo, el trazo horizontal discontinuo en $\alpha =23\ \rm{Np/mm}$ indica que el máximo grabado posible es de $40\ \rm{nm}$, con lo que se elige un grabado inferior de $30\ \rm{nm}$ para tener en cuenta un posible error en el grabado, ya que el fabricante Applied Nanotools, normalmente empleado por el grupo, especifica una tolerancia de grabado de $\pm10\ \rm{nm}$. 
Para $H_g=30\ \rm{nm}$ y $\Lambda=0.45\ \upmu\rm{m}$ la fuerza de radiación conseguida es de $15\ \rm{Np/mm}$, lo cual permite mejorar la especificación inicial y aumentar la dimensión de la rejilla a $150\ \upmu\rm{m}$. Además, es posible reducir aún más la fuerza de radiación si el ciclo de trabajo se optimiza adecuadamente. En la Fig. \ref{fig:GraficasJuntas}(b) se representa un barrido del ciclo de trabajo entre $22\%$ y $77\%$, lo que para $\Lambda=0.45\ \upmu\rm{m}$ garantiza de sobra el tamaño mínimo de detalle de $60\ \rm{nm}$ que especifica Applied Nanotools. Se comprueba que para minimizar la fuerza de radiación es necesario alejarse de un ciclo de trabajo del $50\%$. Se elige así un ciclo de trabajo cercano al límite, del $25\%$, reduciendo la fuerza de radiación a $5.6\ \rm{Np/mm}$. Esta fuerza corresponde a una rejilla de aproximadamente $400\ \upmu\rm m$ y que emite un haz que tiene una distancia de Rayleigh de $4.3\ \rm{cm}$. El ángulo de radiación del campo calculado por el simulador MP es $\theta=-28.2^{\circ}$ medido con respecto al eje vertical $z$, que coincide con el ángulo que arroja FEXEN de $28.15^{\circ}$, tal y como se ilustra en la Fig. \ref{fig:campo}(a), donde se representa la parte real del campo radiado simulada con FEXEN. En la Fig. \ref{fig:campo}(b) se muestra tanto la intensidad de campo radiado en un corte en $z$, que se rige por una exponencial decreciente con una constante de atenuación $2\alpha = 11.2 \rm{Np/mm}$, $\alpha = 5.6 \rm{Np/mm}$, como el diagrama de radiación asociado, con un FWHM ("Full-Width  Half-Maximum") de $0.18^{\circ}$.

Finalmente, se completa el diseño analizando tanto la direccionalidad ofrecida por la estructura como las reflexiones producidas. Para la direccionalidad se obtiene un valor satisfactorio del $58\%$, tanto en el simulador MP como en FEXEN. No obstante, el cálculo de la potencia reflejada no está implementado todavía en el simulador que aquí se presenta, y por ello se realiza con FEXEN, llegando a un valor de $|s_{11}|^2=-34\ \rm{dB}$, despreciable a efectos prácticos.

En resumen, las dimensiones diseñadas para la rejilla de difracción de la Fig. \ref{fig:Grating2D} son: $H_{\rm{BOX}}=2\ \upmu\rm m$, $H_f=190\ \rm{nm}$, $H_g=30\ \rm{nm}$, $\Lambda=0.45\ \upmu\rm m$ y $a=112\ \rm{nm}$, y las características que ofrece son: $\alpha=5.6\ \rm{Np/mm}$, $\eta = 0.58$, $\theta=-28.15^{\circ}$ y $|s_{11}|^2=-34\ \rm{dB}$. La longitud de la rejilla conseguida es de $L_g\approx400\ \upmu\rm m$, radiando un campo con perfil exponencial con una distancia de Rayleigh de $4.3\ \rm{cm}$.


\section{Conclusiones}
En este trabajo se ha desarrollado un simulador electromagnético de rejillas de difracción en guias dieléctricas basado en el método perturbacional. Este se ha aplicado por primera vez y con excelentes resultados al diseño de una rejilla de refracción débil en SOI de longitud $0.4\ \rm{mm}$, capaz de emitir un haz colimado con una distancia de Rayleigh mayor de $4\ \rm{cm}$. La herramienta desarrollada ofrece una solución rápida y completa para diseñar rejillas de difracción, mejorando, hasta en un factor de $\times 16$ los tiempos de simulación cuando se compara con la herramienta FEXEN.


\section*{Agradecimientos}
Este trabajo ha sido financiado por el Ministerio de Universidades, Ciencia e Inovación bajo el proyecto PID2022-139540OB-I00, el proyecto TED2021-130400B-I00/ AEI/10.13039/501100011033/ Unión Europea NextGenerationEU/PRTR, el proyecto PDC2023-145833-I00, la cátedra PERTE-Chip bajo el proyecto TSI-069100-2023-0013 y la Universidad de Málaga.




\bibliography{Bibliografia} 



\bibliographystyle{unsrt}

\end{document}